\documentstyle[preprint,aps]{revtex}
\begin{document}
\draft

\title{{\bf Exact Diagonalization Study of the  \\
One-dimensional Disordered XXZ Model}
} \author{{Karl J. Runge}
} \address{{Lawrence Livermore National Laboratory,\\ P.O. Box 808,\\
Livermore, California 94550}
} \author{{Gergely T. Zimanyi}
} \address{{Department of Physics, University of California,\\ Davis,
California 95616}
} \date{Sub. PRB 11/10/93} \maketitle \begin{abstract}
We investigate the one-dimensional quantum $XXZ$ model in the presence of
diagonal
disorder. Recently the model has been analyzed with
the help of field-theoretical renormalization group methods, and a phase
diagram
has been predicted. We study the model with exact
diagonalization techniques up to chain lengths of 16 sites.
Using finite-size scaling methods we
estimate critical exponents and
the phase diagram and find reasonably good agreement with the
field-theoretical results, namely, that any amount of disorder
destroys the superfluidity for $XXZ$ anisotropy $\Delta$
between $ -1/2 $ and $1$, while the superfluidity persists to
finite disorder strength for $ -1 < \Delta < -1/2$ and
then undergoes a Kosterlitz-Thouless type transition.
\end{abstract}  \narrowtext

\section {\bf Introduction}

The two focal points of many-body physics in recent times are the
study of strongly interacting systems and the exotic phenomena induced by
the presence of disorder. The repeated refinements of the analytic
and numerical tools and the accumulation of experience in these major
areas have set the stage for the quest of understanding
{\it strongly interacting disordered systems}. For fermionic
problems the attack started in the field of disordered
semiconductors\cite{finkelshtein,lee}. On the other hand, bosonic studies
were delayed by the haunting puzzles raised by their classical
counterpart, the spin-glass problem; and by the relative inaccessibility
of such systems experimentally. The past few years brought breakthroughs
in both of these directions. A fairly comprehensive picture of the spin-glass
state has been developed\cite{young}, and elaborate techniques established
the dirty superconductors\cite{goldman,dynes,hebard}, helium in
vycor\cite{reppy,ahlers}, and quantum spin chains\cite{who} as
{\it well controllable} experimental realizations of the model,
posing many challenges for theoretical studies.

We concentrate on the phenomena at zero temperature, as the most profound
differences between the ordered and disordered systems
manifest themselves at that point. At $T=0$ the ordered models undergo a
{\it quantum phase transition}, which occurs as a parameter
of the Hamiltonian is tuned across some critical value.
In this case quantum
fluctuations drive the transition instead of the usual thermal ones.
Ordered  $d$ dimensional quantum systems are equivalent
to a corresponding $d+1$ dimensional classical systems and
as such their critical phenomena are well understood.
Whereas for disordered quantum systems the classical analogues
are much less worked out -- {\it e.g.}, the McCoy-Wu model\cite{mccoywu}
is the single exactly solved model -- and as such their study demands
genuinely new theoretical approaches.
The subject of our paper is the numerical
study of this competition between quantum ordering tendencies
 and the  disruptive effects of
disorder on the example of the one-dimensional disordered $XXZ$ model.

The ordered one-dimensional $XXZ$ model with spin ${1/2}$, described
by the Hamiltonian:

\begin{equation}
H=\sum_{i=1}^L ~(S_{i}^{x}S_{i+1}^{x}+S_{i}^{y}S_{i+1}^{y}+
\Delta ~S_{i}^{z}S_{i+1}^{z})
\label{Ham0}
\end{equation}
is one of the most-studied quantum systems and many of its properties are
well known\cite{affleck,mattis}.
For the values of the anisotropy parameter $-1<\Delta\leq1$ the system
develops quasi-long-range order (quasi-LRO). Expectation values
of the spin-operators vanish, but the spin-spin correlation functions decay
only as a power-law in the ground state. For $\Delta>1$ the excitation
spectrum is gapped and
long range antiferromagnetic order of the Ising-type is established,
and similarly for $\Delta\leq -1$ ferromagnetic Ising long range order sets in.

As it is well known, the spin ${1/2}$ $XXZ$ model is equivalent
to a lattice gas of hard core bosons, which is thought of as an approximate
representation of $^{4}$He\cite{matsuda,papafisher}.
The $z$-component of total magnetization and boson particle number
are related by $M_z + L/2 = N_{\rm b} $.
The above antiferromagnetic phase corresponds to the solid phase
of helium. Ordering in the $XY$ plane maps onto the
superfluid phase.

The disorder will be represented by adding a random magnetic field to the
Hamiltonian:
\begin{equation}
H_{random}=2\sum_{i=1}^L ~h_{i}^z S_{i}^{z} .
\label{Hran}
\end{equation}
where $\langle  h_i^z \rangle = 0$, $ \langle  (h_i^z)^2 \rangle = D$.
In the equivalent hard-core boson problem this corresponds to the inclusion of
a random site-energy.
For other types of disorder, such as random bonds, see
Refs. \cite{doty} and \cite{dagotto}.
The effect of disorder in the Ising regimes is well understood.
According to the Imry-Ma argument the long range order is destroyed
by the addition of a weak random magnetic
field in the $z$-direction, however it
is expected to persist in the presence of a weak random
exchange term\cite{imry}.

The quasi-LRO regime has been first studied by
Giamarchi and Shulz (GS) in the presence of disorder\cite{giamarchi},
who developed a powerful scaling scheme for the problem.
They utilize the Haldane representation for the bosons\cite{haldane},
when writing down the effective action:
\begin{equation}
S= \int dx ~d\tau {\kappa\over 2} [(\partial_{\tau}\Phi)^{2} +
(\partial_{x}\Phi)^{2}] + [\eta (x)\partial_{x}\Phi + \rho (x) e^{i\Phi}+
\rho^{*}(x) e^{-i\Phi}],
\label{Saction}
\end{equation}
Here $\Phi$ is the phase field, representing the bosons and $\eta$ and $\rho$
are the Fourier components of the disorder fields at momenta $k\approx 0$
and $k\approx \pi \rho_{0}$, respectively, where $\rho_{0}$ is the average
density of the bosons.
In the general picture of disordered systems, the
backscattering ({\it i.e.}, the large momentum
component) is driving the localization phenomena and this expectation
is borne out by explicit calculation in the present case as well.
In Eq. (\ref{Saction}) $\kappa$ is a spin-stiffness of the ordered system. It
can
be related to the original parameters by analyzing the Bethe-ansatz solution
of the problem for zero disorder to arrive at\cite{baxter}:
\begin{equation}
\kappa = {1\over 2\pi} \Bigl( 1- {1\over \pi} \cos^{-1}\Delta \Bigr).
\label{Stiffness}
\end{equation}
Upon integrating out the disorder one arrives at an action similar to that of
the sine-Gordon problem:
\begin{eqnarray}
S_{\rm eff}=&& \sum_{\alpha=1}^n \int dx ~d\tau {\kappa\over 2}
[(\partial_{\tau}\Phi_\alpha)^{2} + (\partial_{x}\Phi_\alpha)^{2}] \nonumber\\
&& -D\sum_{\alpha,\beta=1}^n
\int dx ~d\tau ~d\tau' \cos [\Phi_\alpha (x,\tau) - \Phi_\beta (x,\tau') ],
\label{Seff}
\end{eqnarray}
where
the $\alpha$ and $\beta$ sums are over the $n$ replicas.\cite{sorryd}
The infrared singularities are then taken care
of by a renormalization group analysis, which yields:
$$
\partial D/\partial l = (3-1/2\pi\kappa) D,
$$
\begin{equation}
\partial \kappa/\partial l = {1\over 2} D ,
\label{Recur}
\end{equation}
where $l=\ln ~b$, and $b$ is the scale change ratio. For small values
of the randomness and interactions not too strong, so that
$\kappa < \kappa_{c}=1/6\pi$, $D$ renormalizes
towards a line of fixed points at
$D=0$, and thus the quasi-LRO persists in the presence of disorder.
However above that critical value of
$\kappa_{c}$ the disorder becomes a relevant operator, destroying
the ordering tendencies already for arbitrary small values of $D$.
The phase transition is analogous to the Kosterlitz-Thouless (KT) type.
This scaling analysis can be viewed as the quantum-generalization of the Harris
criterion.

Recently Doty and D. Fisher gave an exhaustive study of the phase
diagram\cite{doty}. By utilizing several scaling arguments in different
parameter regimes, they constructed a schematic phase diagram
for the case of weak random $z-$fields.
In particular, the GS phase transition occurs at
$\Delta=-{1/2}$.
For $-1/2 < \Delta$ the disorder is relevant
and in very small amounts it destroys the quasi-LRO.
Nagaosa has also come to this conclusion\cite{nagaosa}.
This result is not surprising for the $\Delta=0$ ($XY$-model) case which
maps exactly onto noninteracting fermions in a disordered
potential\cite{provers}
and is well-known to become localized with infinitesimal disorder.
In the $-1 < \Delta < -1/2$ regime the quasi-LRO is argued
by Doty and Fisher to be stable
against the disorder up to a finite value of $D=D_c$.
The disordered phase was identified as a ``bose-glass" by M.
Fisher {\it et al.}\cite{mfisher}.
The main physical feature of the bose-glass phase is that all of its
low-lying excitations are localized.
Consequently, it has a vanishing superfluid density
and a finite compressibility. In other words, the localization is achieved
{\it not} by the opening of a gap in the spectrum, {\it i.e.} via the Mott
scenario, but
rather by the Anderson mechanism, which localizes the particles by
the interference of their wavefunctions.
In the localized state the
spin-spin correlations decay exponentially with spatial separation.
By integrating the recursion relations Eqs. (\ref{Recur})
 away from the
Kosterlitz-Thouless  critical regime the
following relation is obtained for
the correlation length in the region $-1/2 < \Delta < 1$:
\begin{equation}
\xi \sim D^{-\phi_{s}},
\label{xi}
\end{equation}
where
\begin{equation}
\phi_{s} = (3-1/2\pi \kappa)^{-1}
\label{phis}
\end{equation}
is a crossover exponent.
$\kappa$ is that of the {\it pure} system (Eq. (\ref{Stiffness})) and so
depends only on
the anisotropy $\Delta$.
Eq. (\ref{phis}) is essentially the
the scaling dimension of the Born-scattering amplitude.
On the other hand, in the
$-1<\Delta<-{1/2}$ region
spin-correlations should still decay as a power-law for small
$D>0$, however the exponent
of the spin-spin correlation function
is modified by the presence of disorder.
The KT transition occurs when disorder increases the stiffness
to $\kappa = 1/6\pi$.
In our work we set out to perform an extensive numerical survey
of the above ideas.

\section {\bf Numerical Method}

We use the exact diagonalization method to compute
the properties of the disordered spin $1/2$ $XXZ$ model.
The random fields $h^z_{i}$ in Eq. (\ref{Hran}) are drawn from an independent
and
uniform distribution at each site, such that
$\langle  h_i^z \rangle = 0$, $ \langle  (h_i^z)^2 \rangle = D$.
For each realization of disorder the ground state
$ | \psi_0 \rangle $ is found via an accelerated power
method\cite{runge}.
Once the vector $ | \psi_0 \rangle $ is computed, expectation
values of observables, such as the ground state energy $E$
and the spin-spin correlation functions,
are determined. The chemical potential $\mu$ and compressibility
$K$ are found by taking discrete
derivatives of the ground state energy with respect to the boson
number $N_{\rm b}$:
$\mu = \partial E /\partial {N_{\rm b}}$
and $K^{-1}
  = L \partial^2 E / \partial {N_{\rm b}}^2$.
All results reported here are for the half-filled density ({\it i.e.} $M_z=0$)
case: $N_{\rm b} = L/2$.

The superfluid density - which is related to the helicity modulus - is
computed from the formula\cite{mef}
\begin{equation} \rho_s = { 1 \over L} { {\partial^2 E} \over {\partial
\theta^2} }
\label{rhos} \end{equation}
via finite differencing with respect to $\theta$, where $\theta$
is the angle of a phase twist applied at the boundary.
A boson hopping to the right through the boundaries acquires a phase
$e^{i\theta}$, while one hopping in the opposite direction acquires
$e^{-i\theta}$. One may think of $\rho_s$ as a measure of the
"degree of sensitivity to boundary conditions": in the quasi-LRO superfluid
state phase coherence is long-ranged enough to yield a finite $\rho_s$,
whereas in the localized phase $\rho_s$ drops off exponentially with
system size $L$.

For each $(\Delta, D)$ pair, $300$
to $5,000$ realizations of disorder are used for averaging.
The system sizes we study are $L=4, 6,\ldots,14, 16$, with the
smallest number of realizations for the larger systems.

\section {\bf Analysis of Data}

First we checked the accuracy of our numerical procedure
on the clean system. We calculated the spin-spin correlation function
$\Gamma_{ij}\equiv\langle S_{i}^{x}S_{j}^{x}+S_{i}^{y}S_{j}^{y} \rangle$
which should behave as $r_{ij}^{-\eta}$ where
$\eta=2\pi\kappa$. For a finite-size system one would expect the
correlation at separation $L/2$ should go as  $L^{-\eta}$, and so
one may estimate $\eta$ by:
\begin{equation}
\eta \approx - { {\ln[\Gamma(L'/2)/\Gamma(L/2)]}
 \over {\ln[L'/L]} }
\label{etaest}
\end{equation}
where $L$ and $L'$ are two different system sizes\cite{nightingale}.
Indeed,using this method\cite{barberdg} we find an
exponent within one percent of
the value given by the Bethe-ansatz solution
for nearly all values of anisotropy $\Delta$ (except near $\Delta=1$,
the isotropic HAF point, where logarithmic corrections complicate
our extrapolation scheme).
We also find these estimates for $\eta$ agree closely with that predicted
from the thermodynamical quantities via $2\pi\eta=1/\sqrt{\rho_s K}$.
Results for $\eta$, $\rho_s$, and the compressibility $K$ are shown
in Fig. (1). Note that while $\rho_s$ is nearly constant in the
entire $-1 < \Delta < 1 $ range, $K$ varies strongly, and diverges
at the isotropic ferromagnetic point $\Delta=-1$ (where all bosons
occupation number sectors have the same ground state
energy, thereby making the system infinitely compressible).
The prediction\cite{giamarchi,doty} as to whether the
system's quasi-LRO will be stable or unstable with respect to the addition of
infinitesimal disorder depends only on the pure system
 quantity $\kappa=1/(4\pi^2\sqrt{\rho_s K})$.
As the compressibility increases the superfluid
phase correlations become stronger ({\it i.e.} decay more
slowly) until the point is reached where
a $D_c > 0$ is required to drive $\rho_s$ to zero.

We started the study of the disordered system by computing the
universal scaling function for the superfluid density.
Utilizing the relation between the current-current correlation
function and $\rho_{s}$\cite{girvin} it is straightforward to derive the
finite-size scaling ansatz:
\begin{equation}
\rho_{s}(L, \xi)\sim \xi^{~2-(d+z)} {\tilde {\rho_{s}}}(L/\xi),
\label{fss}
\end{equation}
where ${{\tilde \rho}_{s}}(x)$ is a universal function. As shown by
Giamarchi and Shulz\cite{giamarchi},
the dynamical critical exponent $z$ is one in one dimension,
thus in fact the superfluid density itself is expected to be a universal
function in our case\cite{anotherxy}.

We determined $\rho_{s}$ for roughly 1000 points\cite{oflight}
in the $(L,\Delta,D)$ parameter space.
Typical data for $\rho_s$ is shown in the insets to
Figs. (2a),
(2b), (2c), and (2d),
and later in Figs. (8a) and (8b).
Before delving into the detailed analysis, some qualitative
statements can be made.
In the large disorder regime ({\it i.e.} $\sqrt{D} > 0.30$),
$\rho_s$ vanishes
quickly with $L$, whereas for $D=0$ it clearly extrapolates to a
non-zero value, indicating the presence of a phase-transition.
Thus the system is much more susceptible to the addition of disorder
than is its two-dimensional (2D) analog.
The 2D case
with $\Delta=0$ and Hamiltonian given by the sum of Eqs. (\ref{Ham0}) and
(\ref{Hran}) was considered by Runge\cite{runge}, where it was
found that the critical value of disorder was $\sqrt{D_c} = 1.3$.
We will see below that for the one dimensional (1D) case in the whole range
$-1 < \Delta < 1$, $\sqrt{D_c}$ is never bigger than $0.1$ or so\cite{2dfoot}.
This means for $-1 < \Delta < 1$ the value of the $D$ parameter in the
effective action of Eq. (\ref{Seff}) that destroys quasi-LRO is no larger
than $0.04$.
Such small values of critical disorder, however, are not unexpected: 1) the
superfluid phase is only power-law correlated as opposed to 2D which possesses
true LRO,
and 2) $D_c$ is expected to be zero for $\Delta < -1$ and for
$-1/2 < \Delta$,
so it is not too surprising that $D_c$ cannot become large in the remaining
interval.

\noindent {\bf Power-law regime: $-1/2 < \Delta < 1$}

To our knowledge the first numerical work on the spin $1/2$ $XXZ$
model in the presence of disorder is due to Nagaosa\cite{nagaosa}
who studied $0 \le \Delta$ and utilized a transfer matrix method
based on the Suzuki-Trotter breakup to deduce the finite temperature
properties of a long ($L=200$ and $10$ Trotter time slices) chain.
Nagaosa predicted scaling relations and obtained very good
scaling functions for the superfluid and charge density wave susceptibilities
as $1/kT\equiv\beta \to \infty$. Our work complements his, as we explicitly
have
$\beta=\infty$ and study finite system sizes. We focus
on the quantities  $\rho_s, K$, and $\xi$.
For some properties it is advantageous to have $\beta=\infty$ since
this is where the quantum critical phenomenon is more naturally described,
that is to say, the phase transition occurs in the ground state as a
parameter in the Hamiltonian is varied.

\noindent {\bf 1)}
For $-1/2 < \Delta < 1$ the correlation
length is predicted to depend on the disorder according to
Eqs. (\ref{xi}), that is, $\xi \sim D^{-\phi_s}$.
Therefore as a first method, motivated by this form and by Eq. (\ref{fss}),
we consider $\rho_{s}(L,D)$ as a function of $x\equiv(D-D_{c})~L^{1/\phi_{s}}$,
and then choose the parameters $D_c$ and $\phi_s$ that
collapse the different $\rho_s$ curves best.
This is done via a non-linear ($\chi^2$) least squares fit. Since
the data points almost never have the same value of $x$ as defined above,
spline fits are used to interpolate data points to common $x$ values.
In most cases data from $L=8$ or $10$ to $16$ is used since the smaller
system sizes tend to not be in the scaling regime.
In Figs. (2a, 2b, 2c, and 2d)
examples of the data collapse for $\Delta = 0, 0.5, -0.25, 1.0$ are
presented\cite{morenumerics}.
The data evidently scale well even out to large $D$ where
$\rho_s$ becomes small. For these values of $\Delta$, the optimum
value of the fitting parameter $D_c$ was zero\cite{dcnegfoot}.
In Fig. (3) we
show best fit values of $\phi_{s}$ as circles for the different
anisotropies $\Delta$. For comparison we display Doty and Fisher's
prediction for the exponent $\phi_{s}$.
The data appears consistent with
the results of the scaling analysis\cite{giamarchi,doty}, namely,
that the critical value of the disorder is zero and that the
crossover exponent is given by Eq. (\ref{phis}).
The accurate extraction of $\phi_s$ and $D_c$ is hampered, however,
as the KT region at $\Delta=-1/2$ is approached. This is not too surprising
as we are using small systems to discern a {\it crossover}
between different forms of critical behavior rather than the critical
behavior alone.
To underpin the above results we conducted several further analyses,
discussed in the following.

\noindent {\bf 2)}
An unconstrained extraction of the correlation length can be
performed from the superfluid density data. We plot $\rho_{s}$
against $\ln(L)$ for the different values of the disorder $D$. Then we shift
each curve horizontally to maximize overlap with its 'neighboring' curve
possessing the nearest value of $D$. This process
is repeated for all $D$-curves
to obtain a sequence of shifts, an example is shown in Fig. (4a).
As $\rho_{s}$ is presumed to be a function exclusively of $L/\xi$, the amount
of
shift to maximize the overlap of the curves with $D_{1}$ and $D_{2}$
is given by $\ln \Bigl(\xi(D_{1})/\xi(D_{2})\Bigr)$.
{}From all the shifts one can build up, to within an overall constant,
the function $\xi(D)$. This method was used by Yoshida and
Okamoto in their study of the 1D spin $1/2$ $XXZ$
model in the presence of an alternating bond perturbation (which has
a remarkable degree of similarity to the present disordered
system)\cite{yoshida}.
Once the optimal shifts are performed the scaling function $\tilde \rho_s(x)$
is obtained again.

Our results for $\Delta=0$ and $0.5$
are shown in Figs. (4a) and (4b), and demonstrate
again that the data scale nicely.
The inset to Fig. (4a)
shows the original unshifted data, $\rho_s$ vs. $\ln(L)$.
We show the extracted correlation length $\xi(D)$ for
different $\Delta$ in Fig (5).
Note the potential power of this method to yield information on the
bulk correlation length even when it is much larger then the
system sizes considered.
A log-log plot of $\xi(D)$ against $D$ displays a linear region from
which one  can extract an estimate of $\phi_s$; the results are displayed
as crosses in Fig. (3). The values of $\phi_s$ are
close to, and, unfortunately, not more accurate than those computed
with method 1) of constrained data collapse. This is not surprising
since the two analyses are closely related.

We note in passing that we have also extracted a "system
size dependent" correlation length from an equation similar to
Eq. (\ref{etaest}):
\begin{equation}
\xi_L^{-1}(D) \approx - { {\ln[\Gamma(L'/2)/\Gamma(L/2)]}
 \over {L-L'} }
\label{xiest}
\end{equation}
where $L'$ is a system size close to $L$. We have found
$\xi_L(D)$ does behave as $D^{-\phi_s}$ for a range of
$D$, however, when $\xi_L$ becomes comparable to $L$ it, of course,
deviates from the  $D^{-\phi_s}$ form. The finite-size scaling ansatz
for $\xi_L(D)$ should be $ \xi_L(D)/L = F(\xi(D)/L)$, where
$F(x)$ is a universal function and $\xi(D)$ is the {\it bulk}
correlation length. Hence the shifting method of the previous
paragraph should also be applicable to this data as well.
For other possible definitions of $\xi_L(D)$ see Ref. \cite{barberdg}.

\noindent {\bf 3)}
In our third method we consider $\rho_{s}$
as a function of $d \equiv \sqrt{D}$ for the different values of
system size $L$.
By the symmetry $h_i^z \to -h_i^z$, $\rho_s$ must be an "even"
function of $d$ (note that $d$ rather than $D$ enters directly
into the Hamiltonian). In particular, $\partial \rho_s / \partial d = 0$
at $d=0$ and at $d=\infty$. Therefore, $\partial \rho_s / \partial d$
takes on a maximum that should be in the transition region since
for all $\Delta$ we expect a jump in $\rho_s$ across $d_c$
for the infinite system. Let $d^\ast(L)$ be the point of
maximum slope of $\rho_s(L,d)$. As $L\to\infty$, $d^\ast(L)$ should tend
to the location of the bulk critical point, $d_c = \sqrt{D_c}$.
Finite size scaling, Eq. (\ref{fss}), indicates this maximum should
occur at $d^\ast(L) - d_c \sim 1/L^{1/\phi_s}$ if $d_c \ne 0$,
and $d^\ast(L) \sim 1/L^{1/2\phi_s}$ if $d_c=0$.
Therefore,
we compute $\partial \rho_{s}(L,d)/\partial d $, find the value $d^\ast(L)$
where it is maximal, and then plot $d^{\ast}(L)$ vs. $L^{-1/2\phi_{s}}$.
Fig. (6) shows
such plots for $\Delta =-0.38,-0.25,0.0,0.5,0.9,1.0$.
As can be seen in the figure, all the data is consistent with the
transition occurring at $D=0$ (as $d^\ast(L)>0$ the $\Delta=-0.38$
curve must evidently bend upward),
that is to say, disorder is a relevant
perturbation.

\noindent {\bf 4)}
Finally, we can proceed by accepting that $D_c=0$ and attempt to
extract $\phi_s$ for each $\Delta$ studied.
Consider now $\rho'_s(0) \equiv \partial \rho_s(L,D)/\partial D$ at $D=0$. This
is the initial rate of change of the superfluid density with respect
to the perturbation $D$.
According to Eqs. (\ref{xi}) and (\ref{fss}) this object should behave as
$\sim L^{1/\phi_{s}}$.
This method was introduced by Yoshida and Okomoto\cite{yoshida}.
Note that in general the unlimited growth of
 $\rho'_s(0)$ with $L$ is a further indication that $D_c=0$\cite{sqrtfoot}.
The the slope of  $\rho'_s(0)$ vs $L$ on a log-log plot should
reveal the value of the $\phi_s$ exponent.
Our results are shown in Fig (7)\cite{xyfoot}.
The straight lines in the plots is the prediction of Doty and
Fisher\cite{doty}. The $\phi_s$ are extracted via a linear
least squares fit and are plotted as squares in
Fig. (3). There is very good agreement between our numerical
results and the predictions of Ref. \cite{doty}:
they agree within 4\%, 3\%, and 2\% at
$\Delta=0.0, 0.5$ and $1.0$, respectively.
When $\Delta$ is close to $-1/2$, however, the analysis is less accurate since
the critical phenomenon is evidently crossing over to a different
form.
On the basis of the above four different methods we can declare with
good deal of certainty that the predictions of the scaling
theory are confirmed by our data in the $-1/2 < \Delta < 1$ regime.

\noindent {\bf Kosterlitz-Thouless regime:  $-1 < \Delta < -{1/2}$}

Next we study the $-1 < \Delta < -{1/2}$ region,
where it is predicted\cite{giamarchi,doty} that weak disorder is
not relevant so that there is a finite region
in the parameter space where the quasi-long-range-order survives.
Here one extracts the dependence of the correlation length on the
disorder by fully integrating the renormalization group
Eqs. (\ref{Recur}). In this case
we have to represent that the disorder $D$ itself is strongly modified by
the scaling.
Close to the critical point the integration gives a
Kosterlitz-Thouless-type formula\cite{doty}:
\begin{equation}
\xi (D) \sim \exp({A/\sqrt{D-D_{c}}})
\label{xikt}
\end{equation}
for $D > D_c$.
The Kosterlitz-Thouless (KT) behavior is difficult to extract
even in clean systems, and even more so for disordered ones. Thus
we confine ourselves to show that the data are {\it consistent} with a KT
form, and attempt to estimate the phase boundary.
We analyze the critical behavior with the same methods as above.

\noindent{ \bf 1)}
We first construct the universal scaling function for the superfluid density.
The only difference is that $\xi$ is now given by Eq. (\ref{xikt}),
so we plot $\rho_s$ against $L\exp(-A/\sqrt{D-D_c})$.
and fit $A$ and $D_c$ to minimize $\chi^2$.
The scaling functions found  for $\Delta=-0.8$ and $-0.6$
are shown in Figs. (8a) and
(8b). Good scaling is evidently achieved, however,
the value of $D_c$ is poorly determined. Roughly, $\sqrt{D_c}$
is found to be about $0.1$ or smaller, with an error bar of the
same magnitude. We believe this problem is due to the very small
value of $D_c$, since the finite-size rounding region is actually
a good deal larger than $D_c$ itself. This behavior can be contrasted with
the 2D classical $XY$ model on the square lattice. For that model,
the "relatively large" Kosterlitz-Thouless $T_c$
can be located quite accurately from
small systems. We have performed Monte Carlo simulations
of the 2D classical $XY$ model helicity modulus ($\rho_s$) to
test our methods of locating Kosterlitz-Thouless phase
transitions. Using systems with linear dimension up to $L=16$
and the same finite-size scaling methods applied here (and below)
we could readily locate $T_c$ to
within 4\% of its accurately established value\cite{minnhagen}.
It is unfortunate that the present model requires $L$ much larger than
$16$ for accurate results\cite{classxyfoot}, since this is near the
limit of the exact diagonalization method.

We try
two fitting forms for $\xi(D)$ in addition
to Eq. (\ref{xikt}): (1) $\xi \sim \exp (A/\sqrt{d-d_c})$, where
$d\equiv \sqrt{D}$, and (2) $\xi \sim |D-D_c|^{-\nu}$.
For $D_c \ne 0$ and sufficiently near the transition form (1)
is equivalent to Eq (\ref{xikt}). We observe for $\Delta=-0.8$ and
$-0.6$ the use of form (1) lowers the optimal $\chi^2$ by about
30\% from that we achieve from use of Eq. (\ref{xikt}).
This observation points to the fact that $D_c$ is so small,
higher order terms in $\xi$  play a large role in describing
the data.
Use of form (2), $\xi \sim |D-D_c|^{-\nu}$, yields over a factor of
two {\it increase} in the value of $\chi^2$ relative to that we find with
Eq. (\ref{xikt}). This is somewhat promising because it at least
suggests the mechanism for the transition in $-1 < \Delta < -1/2$
is of the Kosterlitz-Thouless form.
Similarly, in the
region $ 0 \le \Delta < 1$ we find the use of the KT $\xi(D)$ (Eq.
(\ref{xikt}))
fitting form yields $\chi^2$ values 3 to 8 times {\it larger} than that from
the power-law form for $\xi(D)$, thereby bolstering the
belief that the power-law is the correct form for $ -1/2 < \Delta < 1 $.

\noindent{ \bf 2)}
For completeness, we apply method 2) of the previous section involving
$\rho_s$ vs $\ln L$ shifting to obtain the scaling function
to the $-1 < \Delta < -1/2$
data. The scaling function for $\Delta=-0.5$ is shown in
Fig. (9). Once again, although a reasonable scaling
function is obtained, the extraction of $D_c$ values from the
resulting $\xi(D)$ proves difficult. The correlation length
$\xi(D)$ does appear to grow much more rapidly than the
data for $0\le\Delta$, which at least hints at the expected KT
behavior.

\noindent{ \bf 3)}
We apply
the maximum slope method as well. Setting $\xi\sim L$
yields the finite-size scaling prediction
$d^\ast(L) - d_c \sim (1/\ln L)^2$.
Fig. (10) displays the $d^\ast(L)$ data for
$\Delta=-0.95,-0.8,-0.6,-0.5$. All extrapolate to values of
$\sqrt{D_c} < 0.15$, consistent with the values of $D_c$ crudely
estimated with method 1) above. The value nearest the isotropic
ferromagnetic point, $\Delta=-0.95$, is the smallest with
$\sqrt{D_c} \approx 0.03$ and is consistent with the prediction
of Doty and Fisher that $D_c \to 0$ as $\Delta \to -1$ from
above\cite{fmdoty}. However, this method yields its largest value of $D_c$
at $\Delta=-1/2$ which is inconsistent with the renormalization
group prediction that $D_c=0$ at the point where the
pure system  $\kappa=1/6\pi$.
To investigate this effect further, we apply the $d^\ast(L)$ vs. $(1/\ln L)^2$
method to the remainder of the data ({\it i.e.} $-1/2 < \Delta < 1$
where the power-law form for $\xi$ and $D_c=0$ are expected).
The estimated $\sqrt{D_c}$ values are plotted as squares in
Fig. (11).
The error bars for these points are estimated roughly by looking
at the linear least squares fitting error for linear and for
quadratic fits to $d^\ast$ vs $(1/\ln L)^2$, and also by extrapolating
$D^\ast(L)=(d^\ast(L))^2$ with similar forms.
Near $\Delta=0$ this method is consistent again
with $D_c=0$ as was shown previously
using the power-law extrapolation. There is, in fact, enough
curvature in the $d^\ast(L)$ vs. $(1/\ln L)^2$ plots for
$\Delta=-0.25$ and $0.0$ to suggest that the simple extrapolations
we perform here are inadequate.
As mentioned before, in the predicted
crossover regime at $\Delta=-1/2$
severe finite-size errors should be expected.
This is more than likely reflected in our overestimate of the size
of the superfluid region with method 3).
It could be that the almost linear curve viewed over a limited range of
$(1/\ln L)^2$ actually bends over for larger $L$.
So, it is possible that the present extrapolation
method for sufficiently large systems  would yield $D_c=0$ for
$\Delta=-1/2$. At present it is impossible to know for certain whether or not
this will happen.

\noindent{ \bf 4)}
As a fourth method we develop a very different approach that makes
explicit use of the renormalization group results.
{}From our $\rho_s$ and $K$ (compressibility) data for finite systems
in the presence of disorder we can compute the quantity
$\kappa = 1/4\pi^2\sqrt{\rho_{s}K} $.
The scaling analysis suggests that the bose-glass transition takes place
when $\kappa =1/6\pi$, ({\it i.e.} when the power-law decay exponent
$\eta=1/3$).
However in the $\Delta < -{1/2}$ regime
the disorder renormalizes the value of $\kappa$, thus the
pure system Bethe-ansatz-based Eq. (\ref{Stiffness}) cannot be applied.

This method involves finding the
point $D^\ast(L)$ where $\kappa(L,D^\ast)=1/6\pi$, and
extrapolating $D^\ast(L)$ to $L=\infty$.
Finite-size scaling indicates the correction should again go
as $(1/\ln L)^2$. This $\kappa$ extrapolation method has, of
course, $D_c=0$ for $ -1/2 \le \Delta$ guaranteed
(for large enough $L$).
Since the field theoretic arguments\cite{giamarchi,doty}
are compelling\cite{bolfoot}, this method may prove to be the
most accurate one near $\Delta=-1/2$. The results of this
extrapolation method are shown as diamonds in Fig. (11).
Error estimates are performed as above in 3) by examining a
number of extrapolation forms along with the statistical error.
It is promising that at $\Delta=-0.8$ this method gives the
same value of $D_c$ as did the maximum slope method 3).
However, at $\Delta=-0.95$ the $\kappa$-extrapolation method
predicts a somewhat larger value of $D_c$ than method 3), although
they evidently agree within large error bars.
Near $\Delta=-1$ there is a large renormalization of the compressibility
$K$ as disorder is turned on. For example, at $\Delta=-0.95$ the
pure system $K$ is near $9$, whereas in the transition region
it is reduced to around $4$.
This rapid variation introduces a large extrapolation of
$d^\ast(L)$ to the thermodynamic limit.
As an aside, we mention that {\it all} of our numerical
data strongly imply a finite compressibility $K$ at the transition
points in accordance with the general predictions of the transition
to the Bose Glass phase\cite{mfisher}.

\noindent{ \bf 5)}
The final method is very similar to the previous one. It utilizes
the interesting property of the KT recursion relations that
the finite-size corrections to $\kappa$ in Eqs. (\ref{Recur})
{\it at} the transition point are universal\cite{minnhagen}.
This result may be derived by expanding $\kappa$ near the transition
point as $\kappa=1/6\pi + \varepsilon$, then Eqs. (\ref{Recur}) become:
$$
\partial D / \partial l = \alpha \varepsilon D
$$
\begin{equation}
\partial \varepsilon/\partial l = {1\over 2} D ,
\label{NewRecur}
\end{equation}
where $\alpha=18\pi$. These yield $d D / d \varepsilon = 2 \alpha \varepsilon$
which may be integrated to give $ D = \alpha \varepsilon^2 + {\rm const}$.
The critical manifold has ${\rm const}=0$, which upon integrating
from a starting length scale $l_0$  to the system size $l=\ln L$
gives
\begin{equation}
  { 1 \over {\varepsilon(l_0)}}
  -{ 1 \over {\varepsilon(l)}}
= { \alpha \over 2}(l - l_0) \label{epsint}
\end{equation}
or $\epsilon(l) = -2/(\alpha l)$ as $l$ gets large.
Thus at the transition point $D_c$ we expect the finite-size formula
\begin{equation}
\kappa(L,D_c)= {1 \over {6\pi} }
- {1 \over {9\pi \ln L}} + \cdots.
\label{ktfss}
\end{equation}
Note that the presence of the cutoff $l = \ln L$ is (necessarily)
asymptotically independent of the fraction of $L$ one selects, since
$ \ln (aL) = \ln L + \ln a \approx \ln L$.
This method has been used for the 2D classical $XY$ model
to locate $T_c$\cite{minnhagen}, and so we attempt to use it here
on our disordered quantum $XXZ$ model. Fig. (12)
shows the $2\pi [ \kappa(L,D) - 1/6\pi] $ vs
$1/\ln(L)$ data for $\Delta=-0.8$
along with the straight line prediction of Eq. (\ref{ktfss}). A
quadratic curve has been added to suggest the possible
asymptotic behavior ({\it i.e.} critical manifold) of the $D=0.125$ curve.
It is reassuring that this value of $D_c$ is consistent with and
close to the values obtained in methods 3) and 4). These three methods
show that our results for $\Delta=-0.8$ are indeed strongly suggestive of
the Kosterlitz-Thouless scenario.

As a final indication that the superfluid phase persists to
finite disorder $D$, consider
Fig. (13), where we plot the log of the
spin-spin correlation function
$\langle S_{i}^{x}S_{j}^{x}+S_{i}^{y}S_{j}^{y} \rangle$ versus
$\ln(L)$ in the top panel and versus $L$ in the bottom panel.
The data is for $\Delta=-0.8$ and $\sqrt{D}=0.100$ and $0.225$.
In the top panel a power-law decay should yield a straight line,
and, indeed, the $\sqrt{D}=0.100$ data does this, whereas the
$\sqrt{D}=0.225$ curve drops off more quickly, indicating exponential
decay. In the bottom panel the log-linear plot does suggest that the
$\sqrt{D}=0.225$ plot is approaching straight line behavior
({\it i.e.} exponential decay), whereas
the $\sqrt{D}=0.100$ still has significant upward curvature.
These data are consistent with the previous $\rho_s$, $K$
analyses that indicate $D_c$ is around $0.125$ or so.

Before closing we mention that recently a Quantum Monte Carlo study of
the bose-Hubbard model has been performed\cite{us}, which
is expected to exhibit the same critical behavior and generic
phase diagram as the present model.
The existence of superfluidity in the presence of disorder
was well established in that study.
Indeed, in the soft core model of Ref. \cite{us} ({\it i.e.}
with coulomb repulsion $U < \infty$), in general one expects
a larger compressibility $K$ than for the corresponding hard core
$(U=\infty)$ model. Through $\eta \propto 1/\sqrt{\rho_s K}$, this
leads to smaller values of $\eta$ and hence a wider range of
stability of the superfluid phase. Since the model of Ref. \cite{us}
corresponds to $\Delta=0$ ({\it i.e.} no nearest neighbor interaction),
there should exist a certain value of $U$ at which the
$\eta$ of the pure system will be $1/3$, and above this value
of $U$ infinitesimal disorder destroys the superfluidity.

In conclusion we studied the quantum spin $1/2$ $XXZ$ model in one-dimension
with diagonal disorder via exact diagonalization techniques.
By employing different finite-size scaling methods we
mapped out the phase diagram shown in Fig. (11).
We found that weak disorder is relevant in the $-1/2 <\Delta < 1$ regime,
or equivalently the critical value of the disorder is zero.
Our estimate of the
power-law exponent $\phi_s$ was found in agreement with
the predictions of Doty and Fisher\cite{doty}.
For $-1 < \Delta < -1/2$ the results suggest that a small, but
finite disorder is needed to destroy the quasi-LRO.
Thus at small disorder the superfluidity prevails as shown in
 Fig. (11).
Our data in the transition region can be described by the
Kosterlitz-Thouless form.
Therefore, the overall picture emerging from our analysis is in
agreement with the field theoretical and renormalization group predictions
of Refs. \cite{doty}, \cite{giamarchi}, and \cite{nagaosa}.

We acknowledge useful conversations with R.T. Scalettar, G. Batrouni,
D.S. Fisher, T. Giamarchi, and A.P. Young.
KJR was supported by the U.~S. Department of Energy by the Lawrence
Livermore National Laboratory under contract number W-4705-Eng-48.
GTZ was supported by the NSF grant DMR 92-06023.
Computations were performed at the Cornell National Supercomputer
Facility.

\vfill\eject
\section { \bf Figure Captions}

Fig. (1)
Pure system superfluid density $\rho_s$, power-law decay
exponent $\eta$, and compressibility $K$ for the 1D spin $1/2$
$XXZ$ model with $z$-anisotropy $\Delta$. $\Delta=1$ corresponds
to the isotropic Heisenberg antiferromagnet (HAF). Curves are the
exact Bethe-{\it ansatz} solutions for the infinite system.
Symbols denote the extrapolation of our $L \le 16$ exact diagonalization
results. $\rho_s$ and $K$ were obtained from numerical finite
differences, while $\eta$ was extracted from the spin-spin
correlation function $\langle S_{i}^{x}S_{j}^{x}+S_{i}^{y}S_{j}^{y} \rangle$.

Fig. (2)
Finite size scaling function for the superfluid density as a function
of the combination $(D-D_c)L^{1/\phi_s}$ (that is simply related
to $\xi/L$)
for the optimal values of the fitting parameters $D_c$ and
$\phi_s$ (method 1). Insets show the original unscaled $\rho_s$ data.
Anisotropies are:
(a) $\Delta=0.0$ ($XY$ model),
(b) $\Delta=0.5$,
(c) $\Delta=-0.25$, and
(d) $\Delta=1.0$ (HAF).

Fig. (3)
Exponent for the power-law divergence of the correlation
length: $\xi(D) \sim$ $|D-D_c|^{-\phi_s}$. Circles denote
values determined by least squares fitting to $D_c$ and $\phi_s$
to best collapse the $\rho_s$ data (method 1).
Crosses were obtain from the unconstrained $\xi$ determination
(method 2) and have been displaced horizontally by $0.05$ for
clarity.
Squares are extracted from the divergence of $\partial \rho_s /\partial D$
at $D=0$ as $L\to\infty$ (method 4).
The curve is the field theoretic renormalization group prediction
of Ref. \cite{doty} (Eq. (\ref{phis})).

Fig. (4)
$\rho_s$ scaling function from unconstrained correlation length
extraction method (2)\cite{yoshida}.
Inset shows $\rho_s$ vs $\ln L$ data before the shift. Straight
lines connect the data in the inset with common disorder $D$ values.
Anisotropies are:
(a) $\Delta=0.0$,
(b) $\Delta=0.5$.

Fig. (5)
Correlation lengths $\xi(D)$ computed via method 2 ($\rho_s$ vs
$\ln L$ shift method). Note that the extracted $\xi$ is much
bigger than the maximum system size ($L=16$).
The $\Delta$ values are listed in the figure, where one notes
that for fixed $D$, $\xi$ is a monotonically decreasing function
of anisotropy $\Delta$.

Fig. (6)
Finite-size extrapolation of the position of maximum slope
$d^\ast(L)$ of $\rho_s(L,d)$ using the value of exponent
$\phi_s$ given in Eq. (\ref{phis}) (method 3).
Extrapolation of $d^\ast$ to zero implies infinitesimal disorder
destroys the superfluidity, {\it i.e.} disorder is a relevant
perturbation.

Fig. (7)
Log-log plot of the initial slope (with respect to $D$) of $\rho_s$ vs $L$
for a number of anisotropies $\Delta$ (method 4).
Straight lines have slope $1/\phi_s$ (with $\phi_s$ from Eq. (\ref{phis}))
and are the predictions of the field theoretical treatment (Ref. \cite{doty}).

Fig. (8)
$\rho_s$ scaling function based on the assumption of the Kosterlitz-Thouless
(KT) correlation length divergence (Eq. (\ref{xikt})), in the region of
$\Delta$ where the KT form is expected (method 1).
Insets show original unscaled data as in Fig. (2).
Anisotropies are:
(a) $\Delta=-0.8$,
(b) $\Delta=-0.6$.

Fig. (9)
$\Delta=-0.5$ scaling function for $\rho_s$ from the unconstrained
$\xi(D)$ determination (method 2), as in Fig. (4).

Fig. (10)
Extrapolation to $L=\infty$ of the position of maximum
slope
$d^\ast(L)$ of $\rho_s(L,d)$, for anisotropies
$\Delta=-0.5,-0.6,-0.8$, and $-0.95$ in the region
where the Kosterlitz-Thouless transition is expected (method 3).

Fig. (11)
Disorder ($D$) -  Anisotropy ($\Delta$) phase diagram for the
random field, one-dimensional, spin $1/2$ $XXZ$ model.
The system is superfluid with quasi-LRO for the pure
system $D=0$ with $-1 < \Delta \le 1$.
For $-1/2 < \Delta < 1$ it is predicted\cite{giamarchi,doty} that infinitesimal
disorder destroys the quasi-LRO. For $-1 < \Delta < -1/2$
it is predicted that there is a Kosterlitz-Thouless transition
at finite $D_c>0$.
The crosses ($\times$) are our results from method (1) of constrained
$\rho_s$ data collapse and are consistent with $D_c=0$;
the squares ($\kern1pt\vbox{\hrule height 1.2pt\hbox{\vrule width 1.2pt\hskip
3pt \vbox{\vskip 6pt}\hskip 3pt\vrule width 0.6pt}\hrule height
0.6pt}\kern1pt$) are from the maximum slope
method (3) assuming the KT form of the transition;
the diamonds ($\diamond$) are from method (4) involving
the stiffness criterion\cite{giamarchi,doty}
$\kappa(D_c)=1/6\pi$.
The region "SF" indicating the superfluid phase in the presence
of finite disorder is a semi-quantitative estimate of the phase
boundary. It is a parabola constrained to
vanish at $\Delta=-1$ and $-1/2$ and go through our estimate
at $\Delta=-0.8$\cite{fmdoty}.

Fig. (12)
Examination of the "flow" of the superfluid stiffness
$\kappa=1/(4\pi^2 \sqrt{\rho_s K})$ with increasing
system size $L$ (method 5). $\kappa_c\equiv 1/6\pi$.
The straight line is the renormalization group prediction
of the finite-size correction at $D_c$ (Eq. (\ref{ktfss})).
The curved line meeting the data point for $D=0.125$
is drawn to suggest the critical manifold: above
this line the system flows to a localized state ($\kappa \to \infty$,
$\rho_s \to 0$), whereas below it the system flows to a
quasi-LRO/superfluid state with $\rho_s >0$ and power-law
decay exponent $(\eta < 1/3)$ renormalized by disorder.

Fig. (13)
Log-log and log-linear plots of
$\langle S_{i}^{x}S_{j}^{x}+S_{i}^{y}S_{j}^{y} \rangle$
at separation $L/2$ for $\Delta=-0.8$, and $\sqrt{D}=0.100$ and $0.225$.
Top panel straight line behavior suggests the $\sqrt{D}=0.100$
point is in the quasi-LRO superfluid phase, while straight line
behavior in the bottom panel suggests the  $\sqrt{D}=0.225$
point has exponentially decaying correlations, {\it i.e.}
it is in a localized phase.

\vfill\eject

\begin{references}
\bibitem{sorryd} After this manuscript was completed we discovered the
missing factor of two that now appears in Eq. (\ref{Hran}).
This means the value of $D$ in Eq. (\ref{Seff}) is actually four times
larger than the $D$ that is defined after Eq. (\ref{Hran}) and used to refer
to the numerical data in the remainder of the text.
Where it matters, we will make it explicitly
clear when we are referring to the value of $D$ in Eqs. (\ref{Seff})
and (\ref{Recur}).
We regret this inconvenience.
\bibitem{2dfoot} Since there are twice as many bonds per site in
2D than in 1D, perhaps a more pertinent quantity to
compare to is $0.5 \sqrt{D_c} = 0.65$.
\bibitem{morenumerics} Numerical calculations were also performed
at $\Delta=-0.37$, and $0.9$  with similar results.
\bibitem{dcnegfoot} Actually the optimal value was slightly negative,
but such values can be ruled out from the symmetry $h^z_i \to -h^z_i$.
The deviation from zero is reflected in our quoted error bars.
\bibitem{sqrtfoot} However, it cannot rule out a strange non-analyticity
such as $\rho_s(D) = \rho_s(0) - \alpha D^{1/2} + \cdots$
\bibitem{fmdoty}
Doty and Fisher\cite{doty} predict the phase boundary near
the isotropic ferromagnetic point to go as
$D_c^{1/2} \sim |\Delta + 1|^{3/4}$ which for the present
numerical data is indistinguishable from the linear dependence
suggested in Fig. (11).
Near $\Delta=-1/2$ Eqs. (\ref{Recur}) imply $D_c^{1/2} \sim |\Delta + 1/2|$.
\bibitem{bolfoot} Note that the very small values of $D_c$
suggested in the present work help to justify the
the small $D$ expansions used in the renormalization
group treatments.
\end{references}
\end{document}